\documentclass{SCIS2026}
\usepackage{makecell}     
\usepackage{booktabs}     
\usepackage{graphicx}
\usepackage{array}
\usepackage{capt-of}

\begin{document}
\ArticleType{RESEARCH PAPER}
\Year{}
\Month{}
\Vol{}
\No{}
\DOI{}
\ArtNo{}
\ReceiveDate{}
\ReviseDate{}
\AcceptDate{}
\OnlineDate{}
\AuthorMark{Chen S et al.}
\AuthorCitation{Chen S, Zhou J, Cheng Z X, et al.}

\title{Advanced 400G-SerDes Modulations and Equalizers for GPU Cluster in AI Era}{Advanced 400G-SerDes Modulations and Equalizers for GPU Cluster in AI Era}

\author[1~\dag]{Sheng CHEN}{}
\author[1~\dag]{Zhengxiang CHENG}{}
\author[2]{Ji ZHOU}{{zhouji@bitzh.edu.cn}}
\author[3]{Haide WANG}{}
\author[4]{Mengqi GUO}{}
\author[1]{\protect\\ [-2pt]Weiping LIU}{}
\author[2]{Liangchuan LI}{}
\author[2]{Xiangjun XIN}{}

\contributions{~These authors contributed equally to this work.}

\address[1]{Department of Electronic Engineering, Jinan University, Guangzhou 510632, China}
\address[2]{Aerospace and Informatics Domain, Beijing Institute of Technology, Zhuhai 519088, China}
\address[3]{School of Cyber Security, Guangdong Polytechnic Normal University, Guangzhou 510665, China}
\address[4]{School of Information Science and Technology, Shijiazhuang Tiedao University, Shijiazhuang 050043, China}

\abstract{With the rapid development of artificial intelligence, a single graphics processing unit (GPU) is insufficient to meet the computing power requirements. To meet the growing demand for computing power, high-speed GPU cluster interconnects are rapidly evolving from 200 Gbps to 400 Gbps. In this paper, we study high-order modulations and advanced equalization techniques to improve the transmission performance of 400 Gbps interconnects. The theoretical bit error rate (BER) expressions for 4/8-ary pulse amplitude modulation (PAM) and approximate BER expressions for six types of PAM6 generated from 32-ary quadrature amplitude modulation (32QAM) are derived. In the additive white Gaussian noise channel, PAM6 generated by framed-cross 32QAM achieves the best BER performance, PAM6 and PAM8 show approximately 4.2  dB and 7.2 dB peak signal-to-noise ratio (PSNR) penalties compared to PAM4 at the KP4 forward error correction (KP4-FEC) limit, respectively. At a line rate of 440 Gbps (i.e., net rate of 400G), the BER performance of PAM4, PAM6, and PAM8 was evaluated through simulations under channels with different degrees of bandwidth limitation. The results show that, as the bandwidth limitation becomes more severe, the dominant performance factor shifts from noise tolerance to bandwidth requirements. PAM4 requires the lowest PSNR to reach the KP4-FEC limit in mildly bandwidth-limited channels. PAM6 achieves the best performance under moderate bandwidth limitation. PAM8 exhibits a significant advantage in severely bandwidth-limited channels owing to its lower bandwidth requirement and reduced susceptibility to high-frequency attenuation. Moreover, with maximum likelihood sequence estimation, both PAM6 and PAM8 meet the KP4-FEC limit under all evaluated channel conditions, while PAM6 requires a lower PSNR to reach the KP4-FEC limit in most channels, highlighting its potential for application in 400G-SerDes systems.
}

\keywords{400G serializer/deserializer, pulse amplitude modulation, feed-forward equalizer, decision feedback equalizer, maximum likelihood sequence estimation.}

\maketitle

\section{Introduction}
\begin{figure}
\centerline{\includegraphics[width=6.5in]{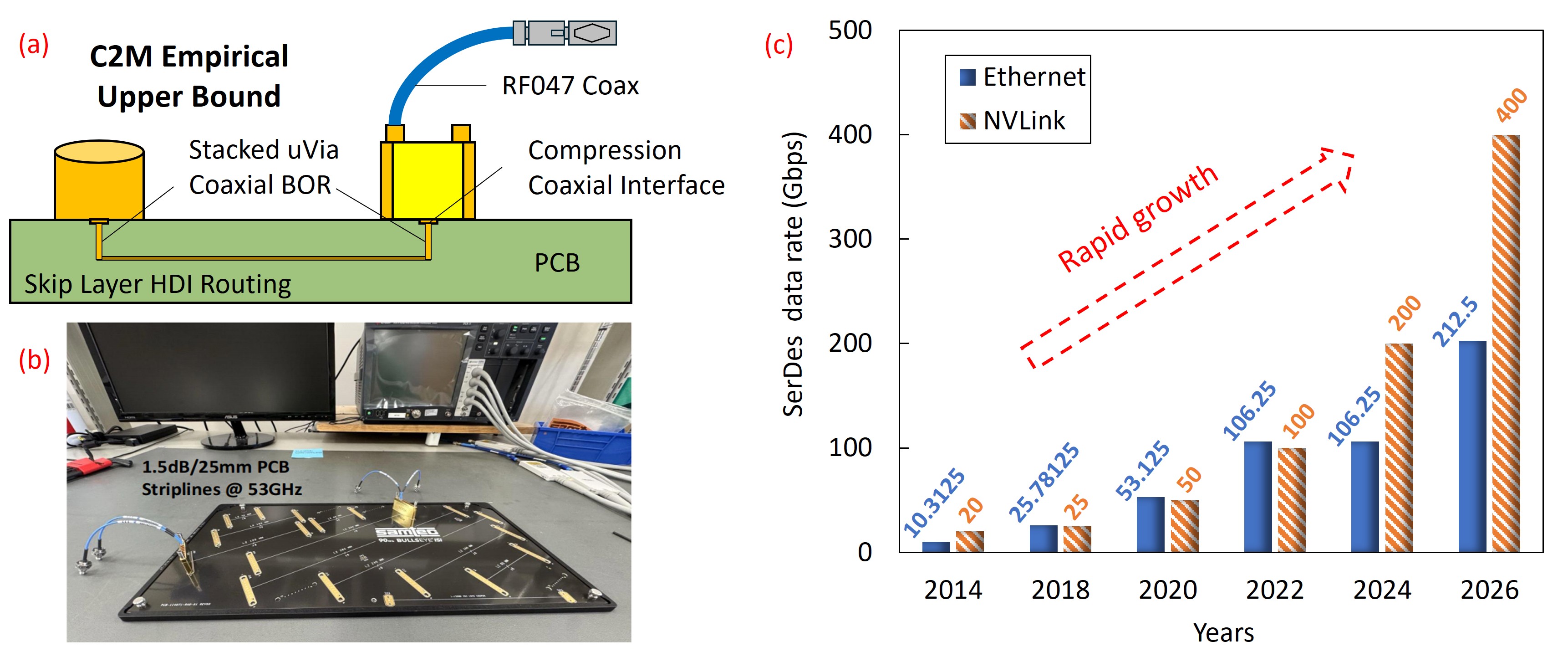}}
\caption{(a) Schematic of a coaxial cable-based C2M interconnect \cite{josephson_e4ai_2025}. (b) Experimental measurement setups for a PCB stripline-based ISI channel emulator \cite{josephson_e4ai_2025}. (c) Evolution of SerDes data rates for Ethernet and NVLink \cite{9999414, NVIDIA2024Blackwell}.
\label{Fig1}}
\end{figure}

With the rapid development of artificial intelligence (AI), a single graphics processing unit (GPU) is insufficient to meet the computing power requirements. The expansion of computing power relies on high-speed interconnect technologies \cite{saber2025physical, liu2024muxflow, nan2026rail}. Fig.~\ref{Fig1} (a) illustrates a coaxial cable-based chip-to-module (C2M) interconnect, in which the printed circuit board (PCB) trace is connected to a coaxial cable through a compression interface \cite{josephson_e4ai_2025}. Fig. 1 (b) shows a PCB stripline-based inter-symbol interference (ISI) channel emulator that provides channels with different losses by varying the stripline length to evaluate modulation formats and equalization techniques \cite{josephson_e4ai_2025}. Fig.~\ref{Fig1} (c) shows the evolution roadmap of serializer/deserializer (SerDes) for Ethernet and NVLink \cite{li2019evaluating, 9999414, NVIDIA2024Blackwell}. In recent years, the data rates of Ethernet and NVLink have increased steadily. Before 2022, Ethernet and NVLink had roughly comparable speeds. In 2024, NVIDIA released NVL72, whose data rate exceeds that of Ethernet. NVIDIA’s sixth-generation NVLink supports per-lane signaling rates of up to 400 Gbps. Thus, 400 Gbps high-speed interconnects are becoming a key technical solution for improving computing power in the AI era.

The limited bandwidth-induced filtering is a main challenge for the 400 Gbps interconnects \cite{Li2026DesignCon}. Therefore, choosing the appropriate modulation format and equalization techniques is of vital importance for 400 Gbps interconnects. The 4-ary pulse amplitude modulation (PAM) has been extensively employed in interconnect systems operating at data rates from 50Gbps to 200Gbps \cite{wang202152}. However, it may not be sufficient to meet transmission performance requirements for single-lane 400Gbps interconnects, prompting the study of PAM6 and PAM8 with higher spectral efficiency (SE) \cite{chun2022pam, khairi20221, hecht2022pam}. In addition, the advanced equalization technologies are required to mitigate limited bandwidth-induced filtering, such as feed-forward equalizer (FFE), decision feedback equalizer (DFE), and maximum likelihood sequence estimation (MLSE).

In this paper, we investigate high-order modulations and advanced equalization techniques to improve the transmission performance of 400 Gbps interconnects. The theoretical bit error rate (BER) expressions for PAM4 and PAM8 are derived, with a particular focus on approximate BER expressions for six types of PAM6 generated from 32-ary quadrature amplitude modulation (32QAM). Three major equalization techniques are analyzed and compared in a 400G-SerDes system under channels with different degrees of bandwidth limitation, including FFE, DFE, and MLSE. The main contributions of this paper are as follows:
\begin{itemize}
\item Approximate BER expressions for six types of PAM6 generated from 32QAM are derived and analyzed, which show that PAM6 based on framed-cross 32QAM achieves the best BER performance.
\item After MLSE equalization, PAM6 achieves the lowest required peak signal-to-noise ratio (PSNR) at the KP4 forward error correction (KP4-FEC) limit under most of the evaluated channels, outperforming PAM4 and PAM8.   
\end{itemize} 

The remainder of the paper is organized as follows. In Section \ref{secII}, the theoretical BER expressions for PAM4 and PAM8 and approximate BER expressions for six types of PAM6 generated from 32QAM are derived, and the advanced equalization techniques are analyzed. Section \ref{secIII} demonstrates the simulation setups of the \mbox{400G-SerDes} system. In Section \ref{secIV}, the simulation results and analysis of the 400G-SerDes system are given. Finally, the paper is concluded in Section \ref{secV}.

\section{Principles of Modulations and Equalizers}
\label{secII}
In this section, the theoretical BER expressions for PAM4 and PAM8 and approximate BER expressions for six types of PAM6 generated from 32QAM under an additive white Gaussian noise (AWGN) channel are derived to quantify the performance penalty caused by noise. Subsequently, the principles of FFE, DFE, and MLSE for 400~Gbps high-speed interconnects are analyzed.

\subsection{Principle of Modulations}
PAM4 has been widely adopted for 200 Gbps high-speed interconnects. However, scaling the data rate to 400 Gbps doubles the baud rate required by PAM4, making the channel bandwidth limitation more severe. Compared with PAM6 and PAM8, PAM4 provides a larger minimum spacing between adjacent signal levels and thus exhibits greater tolerance to noise. In contrast, PAM6 and PAM8 offer higher spectral efficiencies, thereby reducing the required baud rate and relaxing the channel bandwidth requirement \cite{berikaa2023tfln, pittala2025448, ostrovskis2025448, zhou2025enabling}. Therefore, PAM4, PAM6, and PAM8 are all promising candidate modulation formats for 400 Gbps high-speed interconnects. PAM4 maps bits to four signal levels according to the Gray mapping: $(\pm 1, \pm 3)$. However, owing to its relatively high Nyquist frequency, severe channel filtering poses significant challenges. In the peak-power constraint system, the theoretical BER of PAM4 with the PSNR metric is given by 
\begin{equation}
\begin{aligned}
P_{e, \text{PAM4}} & =\frac{3}{4} Q\left(\sqrt{\frac{P S N R}{9}}\right)+\frac{1}{2} Q\left(3 \sqrt{\frac{P S N R}{9}}\right) -\frac{1}{4} Q\left(5 \sqrt{\frac{P S N R}{9}}\right)
\end{aligned}
\end{equation}
where the Q-function is defined as
\begin{equation}
Q(x) = \frac{1}{\sqrt{2\pi}} \int_{x}^{+\infty} e^{-\frac{t^2}{2}} \, dt.
\end{equation}

\begin{figure}[t]
\centerline{\includegraphics[width=6.5in]{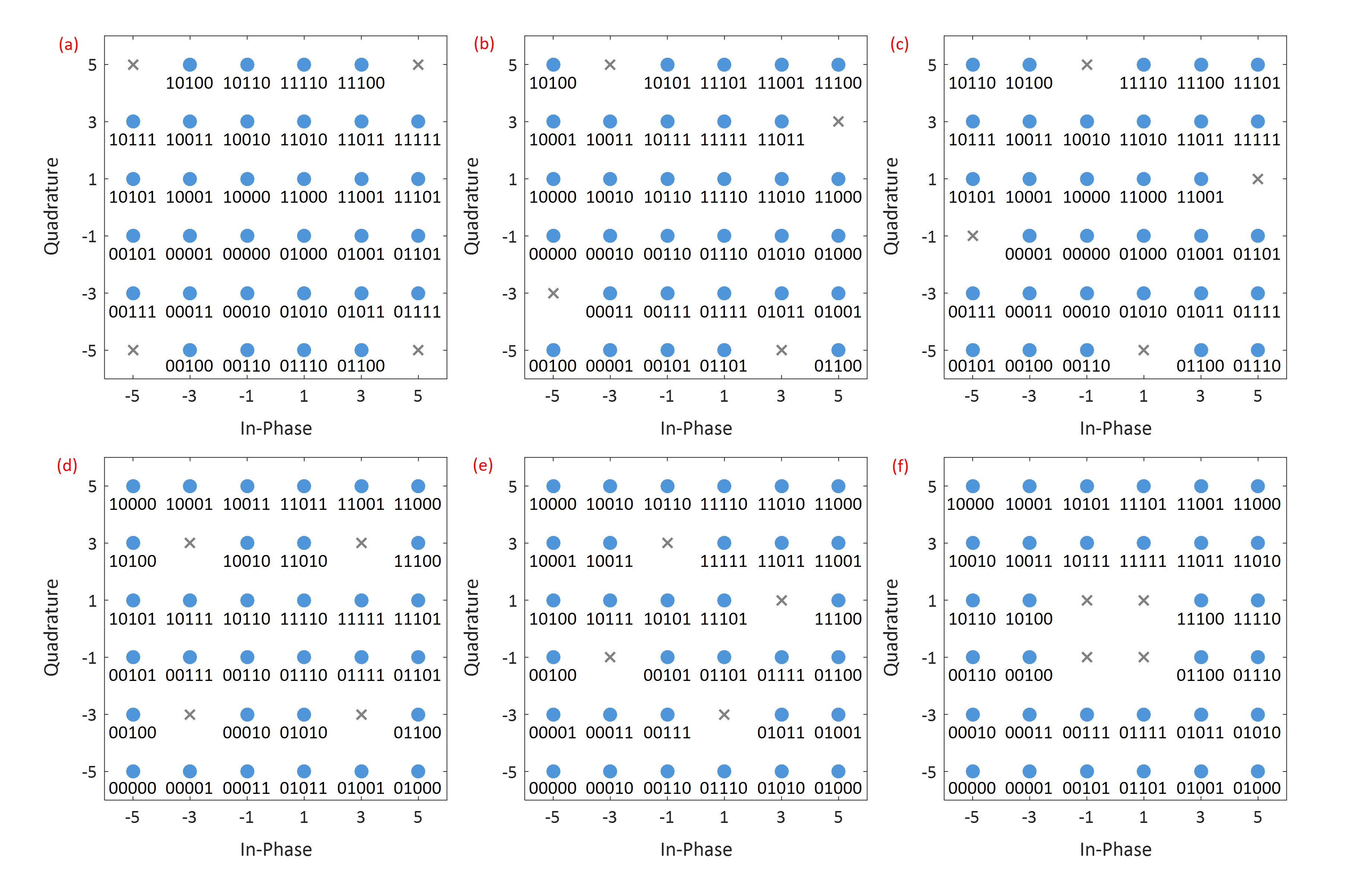}}
\caption{Six types of 32QAM constellation templates for PAM6 generation. (a) Type 1 (cross 32QAM), (b) Type 2, (c) Type 3, (d) Type 4 (framed-cross 32QAM), (e) Type 5, (f) Type 6, respectively. x denotes the discarded constellation point.
\label{Fig2}}
\end{figure}

PAM6 is generated based on 32QAM templates, which maps bits to six discrete levels: $(\pm 1, \pm 3, \pm 5)$ \cite{stojanovic2018210, prinz2022comparison, villenas2025new}. The in-phase (I) and quadrature (Q) components of one complex-value 32QAM symbol form a pair of real-value PAM6 symbols \cite{vitthaladevuni2005exact, cho2002general, smith1975odd}. Thus, each pair of PAM6 symbols represents five bits, and its SE lies between those of PAM4 and PAM8. The 32QAM template is derived from the square 36QAM template by removing four constellation points. Six types of 32QAM constellation templates in Fig.~\ref{Fig2} for PAM6 generation are proposed and analyzed, including (a) Type 1 (cross 32QAM), (b) Type 2, (c) Type 3, (d) Type 4 (framed-cross 32QAM) \cite{prinz2021pam}, (e) Type 5, (f) Type 6, respectively. Among these six types of 32QAM templates, only the framed-cross 32QAM complies with the Gray mapping, which is derived from 36QAM template by discarding four constellation points, i.e., $(-3, -3)$, $(-3, 3)$, $(3, -3)$, and $(3, 3)$. The cross 32QAM is obtained from the 36QAM template by removing its four corner constellation points, i.e., $(-5,-5)$, $(-5,5)$, $(5,-5)$, and $(5,5)$. The approximate BER expression for PAM6 based on the Type $k$ 32QAM can be expressed as
\begin{equation}
\begin{aligned}
P_{e,k} \approx \sum_{i=1}^9 X_{k, i} \cdot Q\left(i \sqrt{\frac{P S N R}{25}}\right) -
\left[\sum_{i=1}^9 Y_{k, i} \cdot Q\left(i \sqrt{\frac{P S N R}{25}}\right)\right] \cdot\left[\sum_{i=1}^9 Z_{k, i} \cdot Q\left(i \sqrt{\frac{P S N R}{25}}\right)\right]
\end{aligned}
\label{eq3}
\end{equation}where the coefficients $X_{k,i}$, $Y_{k,i}$, and $Z_{k,i}$ can be obtained by referring to Table \ref{Tab2}. The detailed derivation of the approximate BER expression for PAM6 is provided in Appendix A.

PAM8 maps bits to eight discrete levels: $(\pm 1, \pm 3, \pm 5, \pm 7)$. With a lower Nyquist frequency, it significantly alleviates the severe high-frequency channel loss compared with PAM4. In addition, each symbol of PAM8 carries 3 bits of information, which enables it to improve SE and achieve higher data rates under bandwidth-limited conditions. As the level spacing decreases, its noise immunity in AWGN channels is significantly degraded. When compared with PAM4 and PAM6, the MLSE equalization complexity of PAM8 is found to increase exponentially. The theoretical BER for PAM8 can be expressed as
\begin{equation}
\begin{aligned}
P_{e, \mathrm{PAM} 8} & =\frac{7}{12} Q\left(\sqrt{\frac{P S N R}{49}}\right)+\frac{1}{2} Q\left(3 \sqrt{\frac{P S N R}{49}}\right)  
 -\frac{1}{12} Q\left(5 \sqrt{\frac{P S N R}{49}}\right) \\&+\frac{1}{12} Q\left(9 \sqrt{\frac{P S N R}{49}}\right) -\frac{1}{12} Q\left(13 \sqrt{\frac{P S N R}{49}}\right).
\end{aligned}
\end{equation}

\begin{figure}[t]
\centerline{\includegraphics[width=6in]{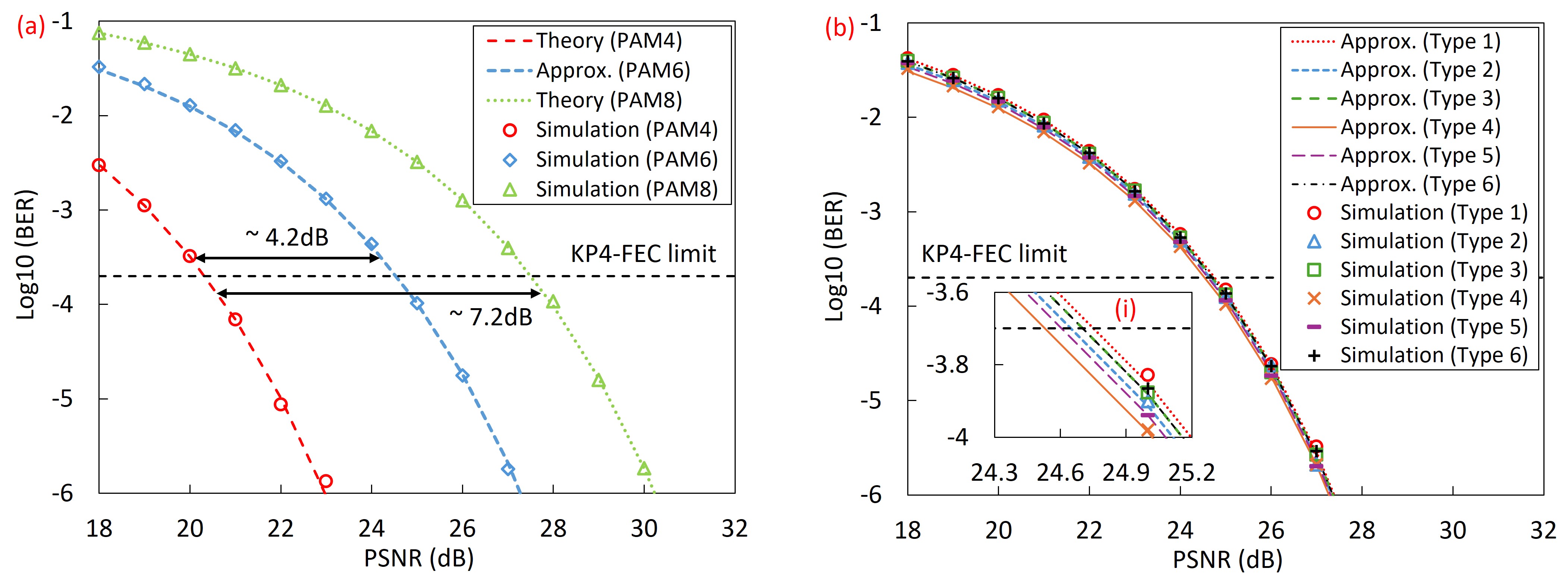}}
\caption{(a) Theoretical and simulated BER performances of PAM4 and PAM8, and approximate and simulated BER performances of PAM6 based on framed-cross 32QAM, using the AWGN channel. (b) Approximate and simulated BER performances of six types of PAM6 based on 32QAM templates. Inset (i) shows the zoom around the KP4-FEC limit.
\label{Fig3}}
\end{figure}

The theoretical and simulated BER performances of PAM4 and PAM8, and the approximate and simulated BER performances of PAM6 based on framed-cross 32QAM, in the AWGN channel are shown in Fig.~\ref{Fig3}(a). The theoretical BER curves of PAM4 and PAM8 and the approximate BER curve of PAM6 agree well with their corresponding simulation results. BER performance of PAM4 serves as a benchmark, while there is about a 4.2~dB and 7.2 dB PSNR penalty in that of PAM6 (Type 4) and PAM8 at the KP4-FEC limit (i.e., BER @ $2\times10^{-4}$), respectively. The approximate and simulated BER performances of six types of PAM6 in the AWGN channel are shown in Fig.~\ref{Fig3}(b). The approximate BER curves closely follow the simulation results over the evaluated PSNR range. Both the approximate analysis and simulation results consistently show that PAM6 generated from framed-cross 32QAM achieves the lowest BER, whereas PAM6 generated from cross 32QAM exhibits the highest BER. At the KP4-FEC limit, PAM6 generated from framed-cross 32QAM requires approximately 0.2 dB less PSNR than PAM6 generated from cross 32QAM.

\begin{figure}[t]
\centerline{\includegraphics[width=6in]{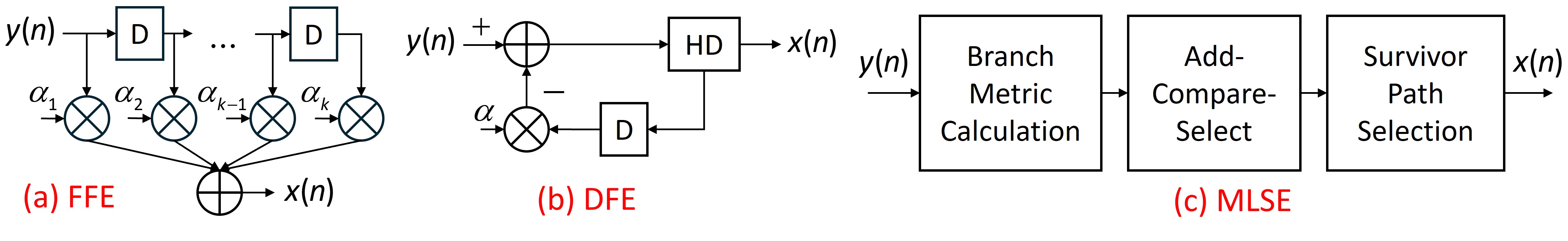}}
\caption{Block diagrams of (a) FFE, (b) DFE, and (c) MLSE for a 400G-SerDes system.
\label{Fig4}}
\end{figure}

\subsection{Principle of Equalizers}
To deal with the filtering caused by the limited bandwidth of the 400G-SerDes system, digital equalizers are employed, including the FFE, DFE, and MLSE \cite{zou2024low, zhou2023burst}. The block diagrams of (a) FFE, (b) DFE, and (c) MLSE are shown in Fig.~\ref{Fig4}, respectively. FFE is a core linear equalization technique that is based on a transversal filter. The input signal $y(n)$ passes through multiple delay units successively. At each delay tap, the signal is multiplied by a weighting coefficient, and all the products are summed to generate the output signal $x(n)$. The DFE eliminates ISI by subtracting the previous hard-decision signals , which are processed via hard decision (HD) and delay units and weighted by the coefficient $\alpha$, from the received signal $y(n)$ \cite{kim201521, park2025low, ok2025critical}. However, since the feedback loop fully depends on the decision output, any decision error propagates through the loop and continuously degrades the recovery of the subsequent signal $x(n)$. Thus, it causes error propagation. MLSE is a theoretically optimal equalization algorithm that determines the most likely transmitted sequence by comparing the received sequence with all possible candidate sequences. It typically employs the Viterbi algorithm to perform dynamic programming over a trellis, thereby approaching the performance of optimum sequence detection\cite{meybodi2022design}. However, the computational complexity of MLSE increases as the modulation order increases.

\section{400G-SerDes Simulation Setups} \label{secIII}
\begin{figure}[t]
\centerline{\includegraphics[width=6.5in]{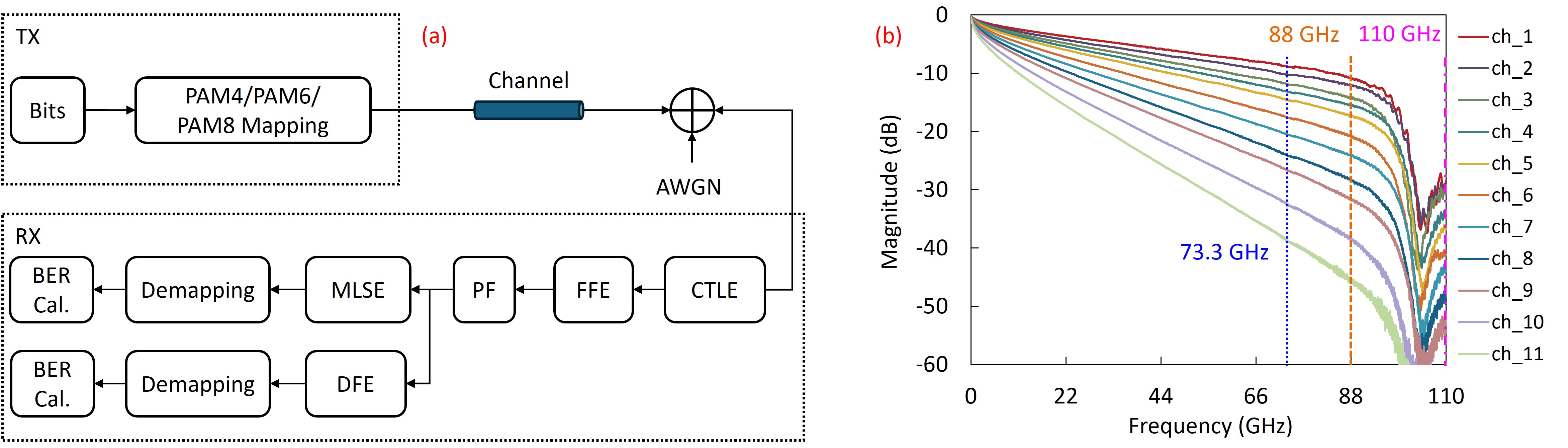}}
\caption{(a) Simulation setups for the 400G-SerDes system. (b) Channel magnitude-frequency response for channel 1-11 \cite{josephson_e4ai_2025}.
\label{Fig5}}
\end{figure}

The simulation setups is shown in Fig.~\ref{Fig5}(a). The transmitter generates a bitstream that is mapped onto PAM4/6/8. Prior to channel transmission, the signal amplitude is limited to 1.2 V peak-to-peak. Eleven measured channel responses from the PCB stripline-based ISI channel emulator in Fig.~\ref{Fig1}(b) are used \cite{josephson_e4ai_2025}. Each channel has 152 mm cables at both ends, and the PCB stripline lengths for channels 1-11 are 40, 60, 80, 100, 120, 160, 200, 240, 280, 360, and 442 mm, respectively. Their magnitude-frequency responses are shown in Fig.~\ref{Fig5}(b), with channel insertion loss increasing as the stripline length increases. The Nyquist frequencies of 440 Gbps PAM4/6/8 are 110 GHz, 88~GHz, and 73.3 GHz, respectively. AWGN is introduced to the signal with a metric of PSNR following the channel.

\begin{table}[!b]
\caption{Parameter settings for the three modulation formats in the representative low- and high-loss channels.}
\label{Tab1}
\centering
\small
\renewcommand{\arraystretch}{1.3}
\setlength{\tabcolsep}{6pt}
\begin{tabular}{l c c c c c c}
\toprule
Channel
& \multicolumn{3}{c}{Channel 1}
& \multicolumn{3}{c}{Channel 11} \\
\cmidrule(lr){2-4}\cmidrule(lr){5-7}
Modulation
& PAM4 & PAM6 & PAM8
& PAM4 & PAM6 & PAM8 \\
\midrule
Symbol Rate ($f$), Gbaud
& 220 & 176 & 146.7
& 220 & 176 & 146.7 \\

DC Gain ($g_{DC}$), dB
& -3 & -1 & 0 
& -20 &  -20 & -18 \\

DC Gain2 ($g_{DC2}$), dB
& -2 & -1 & 0
& -3 & -3 &  -2\\

Low-Frequency Zero ($f_{zm}$), Hz
& $f/160$ & $f/128$ & $f/107$
& $f/160$ & $f/128$ & $f/107$ \\

Low-Frequency Pole ($f_{pm}$), Hz
& $f/160$ & $f/128$ & $f/107$
& $f/160$ & $f/128$ & $f/107$\\

Zero Frequency ($f_z$), Hz
& $f/2.5$ & $f/2.5$ & $f/2.5$
& $f/2.5$ & $f/2.5$ & $f/2.5$ \\

Pole Frequencies ($f_{p1},f_{p2}$), Hz
& $f/2.5$, $f$ & $f/2.5$, $f$ & $f/2.5$, $f$
& $f/2.5$, $f$ & $f/2.5$, $f$ & $f/2.5$, $f$ \\

FFE Tap Number 
& $29$ & $13$ & $13$
& $29$ & $17$ & $13$ \\

Post-Filter Coefficient ($\alpha$)
& $0.9$ & $0.4$ & $0.3$
& $0.9$ & $0.9$ & $0.7$ \\
\bottomrule
\end{tabular}
\end{table}

At the receiver, the signal is first processed by a continuous-time linear equalizer (CTLE) to compensate for the high-frequency attenuation introduced during channel propagation and thereby mitigate ISI \cite{zhang202350gb}. The transfer function of the CTLE is given by
\begin{equation}
\begin{aligned}
H_{\mathrm{CTLE}}(f)=\frac{\left(10^{\frac{g_{D C}}{20}}+j \frac{f}{f_z}\right)\left(10^{\frac{g_{D C 2}}{20}}+j \frac{f}{f_{zm}}\right)}{\left(1+j \frac{f}{f_{p_1}}\right)\left(1+j \frac{f}{f_{p_2}}\right)\left(1+j \frac{f}{f_{pm}}\right)}.
\end{aligned}
\end{equation}
The simulation parameters of CTLE are listed in Table \ref{Tab1}. The symbol rates of PAM4/6/8 are 220 Gbaud, 176~Gbaud, and 146.7 Gbaud, respectively. Higher-order PAM reduces the symbol rate and channel bandwidth requirements, but suffers from lower noise tolerance and higher equalization complexity. The values of $g_{DC}$ and $g_{DC2}$ are listed in Table \ref{Tab1}. This configuration provides the necessary low-frequency attenuation to facilitate relatively high-frequency peaking for channel equalization. The signal is then processed by an FFE to mitigate ISI, with the number of taps specified in Table \ref{Tab1}. Subsequently, the signal is processed by a $(1+\alpha D)$ post-filter (PF) to whiten the colored noise introduced by the FFE, where the value of $\alpha$ is listed in Table \ref{Tab1}. The signal is equalized using either a 1-tap DFE or MLSE. Finally, the symbols are demapped into bits, and the BER is calculated. 

\begin{figure}[t]
\centerline{\includegraphics[width=6in]{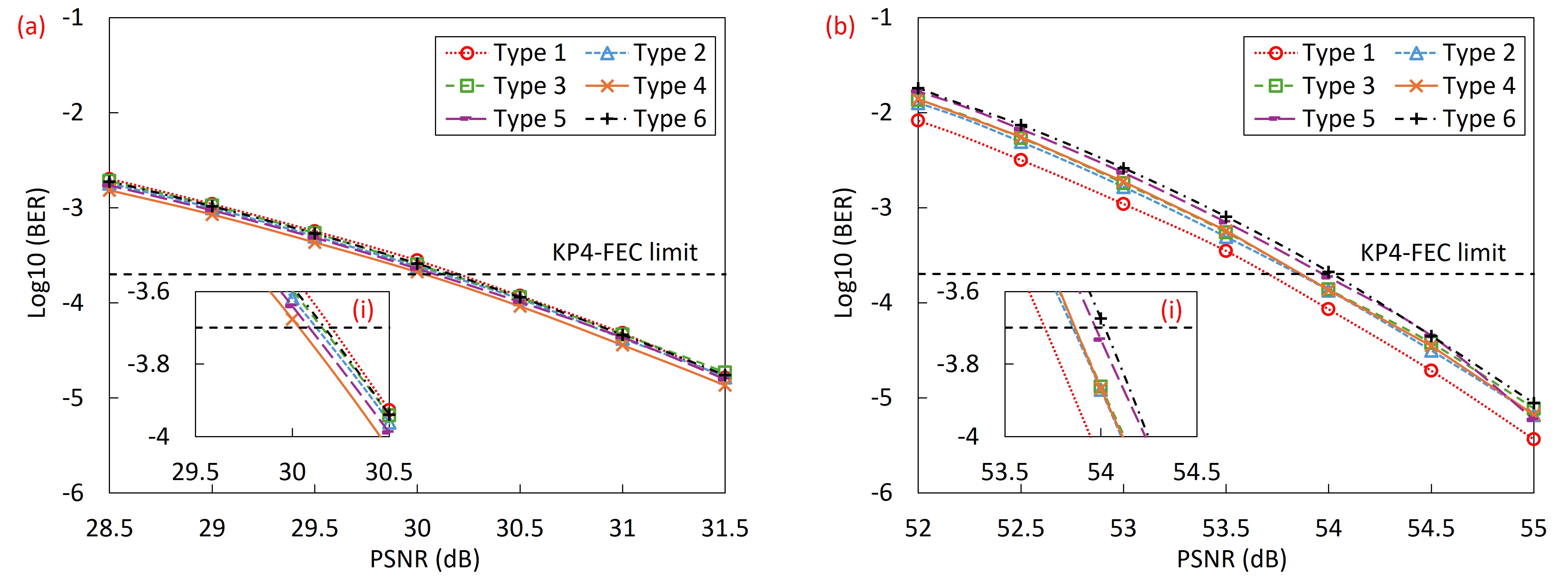}}
\caption{BER performance of MLSE for PAM6 based on six types of 32QAM templates using channel 1 and channel 11. (a) Channel 1, (b) Channel 11, respectively. Inset (i) is the zoom around the KP4-FEC limit.
\label{Fig6}}
\end{figure}

\begin{figure*}[t]
\centerline{\includegraphics[width=6in]{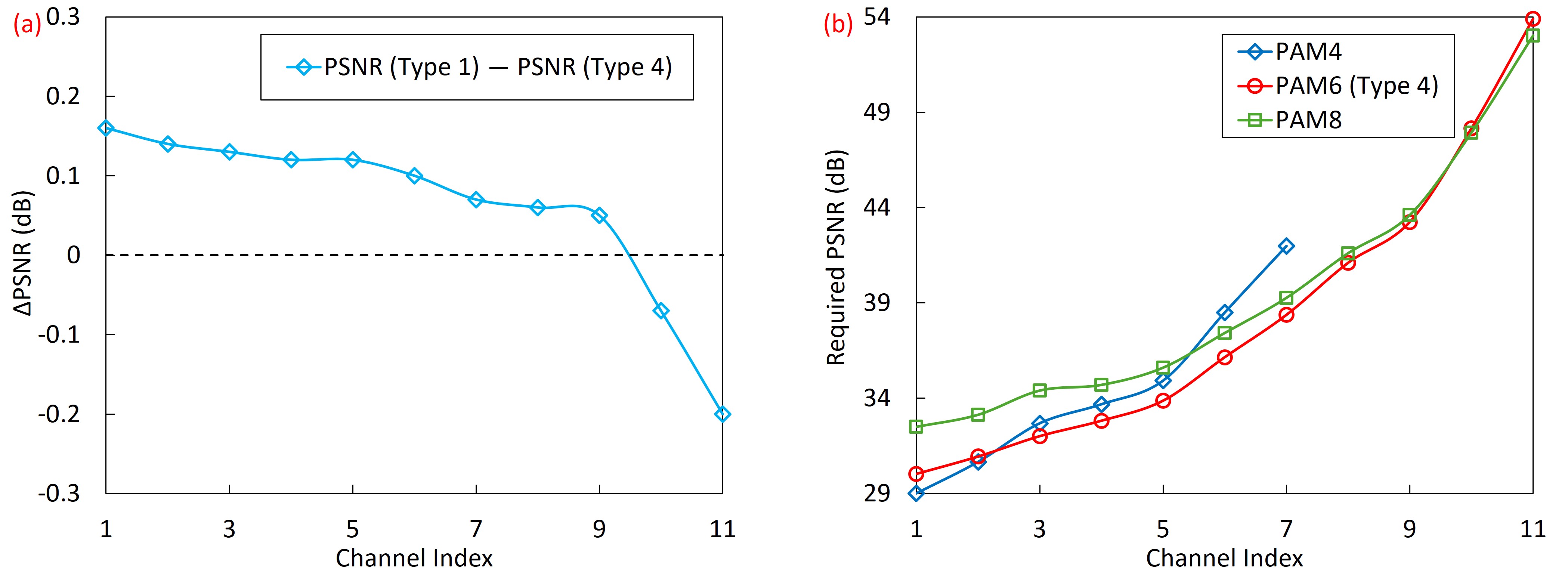}}
\caption{Required PSNR at the KP4-FEC limit after MLSE equalization under channels 1-11. (a) PSNR difference between PAM6 (Type 1) and PAM6 (Type~4), defined as $\Delta\mathrm{PSNR}
= \mathrm{PSNR} (\mathrm{Type}\ 1)
- \mathrm{PSNR} (\mathrm{Type}\ 4)$, (b) required PSNR for PAM4, PAM6 (Type 4), and PAM8. The black dashed line in (a) indicates equal required PSNR for Type 1 and Type 4.
\label{Fig7}}
\end{figure*}

\section{400G-SerDes Simulation Results and Discussion} \label{secIV}
Fig.~\ref{Fig6} (a)-(b) shows the BER performance of PAM6 based on the six types of 32QAM templates using the channel 1 and channel 11, respectively. Insets are the zoom around the KP4-FEC limit. For the 440 Gbps PAM6 system, channel 11 exhibits higher insertion loss than channel 1. In channel 1, the PAM6 based on framed-cross 32QAM achieves the best performance among the six types of PAM6 generated from 32QAM templates. The PAM6 based on framed-cross 32QAM reaches the KP4-FEC limit at a PSNR of approximately 30.03 dB after using MLSE, providing a performance gain of about 0.16 dB over cross 32QAM with the worst BER. For the channel 11, the PAM6 based on cross 32QAM is identified as the optimal mapping scheme for PAM6. Using MLSE, it requires a PSNR of approximately 53.7 dB to reach the KP4-FEC limit, representing an improvement of about 0.3 dB over the Type 6 32QAM scheme. As the in-band insertion loss increases, the performance advantage of PAM6 based on cross 32QAM becomes more pronounced. This advantage primarily stems from removing the \((\pm5,\pm5)\) constellation points. Since these points exhibit more pronounced tailing under high insertion loss, their removal effectively mitigates ISI. Therefore, PAM6 based on cross 32QAM with FFE and MLSE achieves the best BER performance among the six types of PAM6 generated from 32QAM templates under channel 11.

To assess the effect of channel insertion loss on the performance of PAM6 Type 1 and Type 4, Fig.~\ref{Fig7}(a) compares their required PSNRs at the KP4-FEC limit under channels 1-11. A positive $\Delta\mathrm{PSNR}$ indicates that Type 4 requires a lower PSNR, whereas a negative value indicates that Type 1 requires a lower PSNR. In channels with low insertion loss, the relatively weak impact of ISI allows PAM6 (Type 4) to retain its BER advantage observed in the AWGN channel, resulting in a lower required PSNR. As channel insertion loss increases, $\Delta\mathrm{PSNR}$ decreases and becomes negative, indicating that PAM6 (Type 1) requires a lower PSNR than PAM6 (Type 4) to reach the KP4-FEC limit in channels with higher insertion loss. This trend is attributed to the removal of the outer-corner constellation points in Type 1, as these points exhibit more pronounced tailing and their removal helps mitigate ISI. Fig.~\ref{Fig7}(b) compares the required PSNR of PAM4, PAM6 (Type 4), and PAM8 at the KP4-FEC limit. In the low-loss channels, PAM4 requires the lowest PSNR because of its superior noise tolerance. As the insertion loss increases, its required PSNR rises rapidly owing to the larger bandwidth demand. In channels with moderate insertion loss, PAM6 combines a lower symbol rate than PAM4 with better noise tolerance than PAM8, resulting in the lowest required PSNR among the three modulation formats. In channels with the most severe bandwidth limitation, the benefit of PAM8’s lower symbol rate in reducing ISI dominates its performance, giving it an advantage over PAM4 and PAM6 despite its lower noise tolerance. Across most of the evaluated channels, PAM6 achieves the most favorable transmission performance by balancing noise tolerance and bandwidth efficiency. Therefore, PAM6 is a promising modulation format for 400G-SerDes interconnects.

\begin{figure*}[t]
\centerline{\includegraphics[width=6.5in]{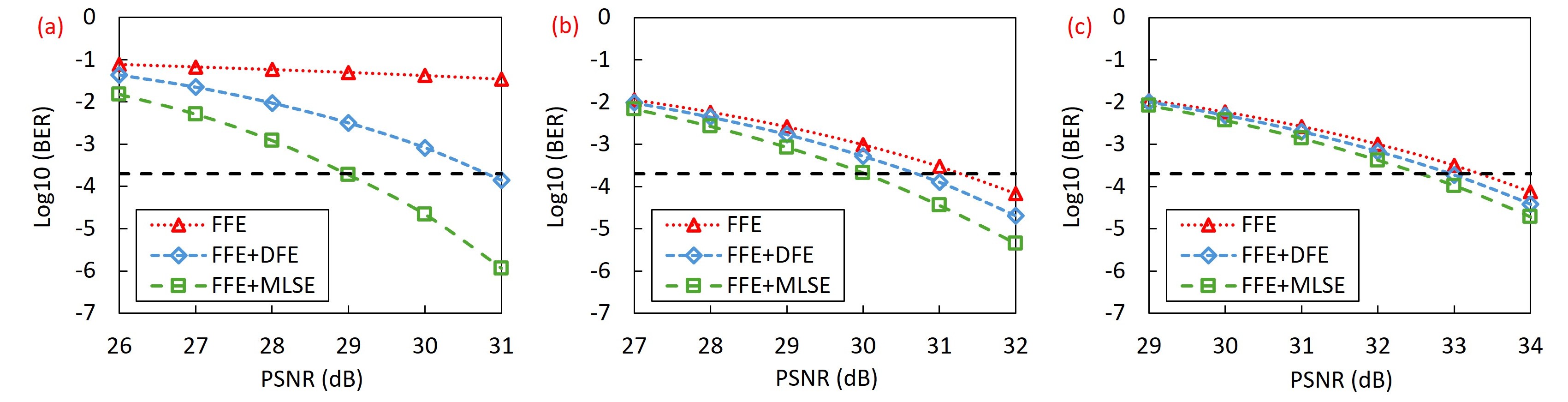}}
\caption{BER performance of (a) PAM4, (b) PAM6 based on framed-cross 32QAM, and (c) PAM8 signals equalized by FFE, DFE, and MLSE using the channel 1. The black dashed line represents the KP4-FEC limit.
\label{Fig8}}
\end{figure*}

\begin{figure*}[t]
\centerline{\includegraphics[width=6.5in]{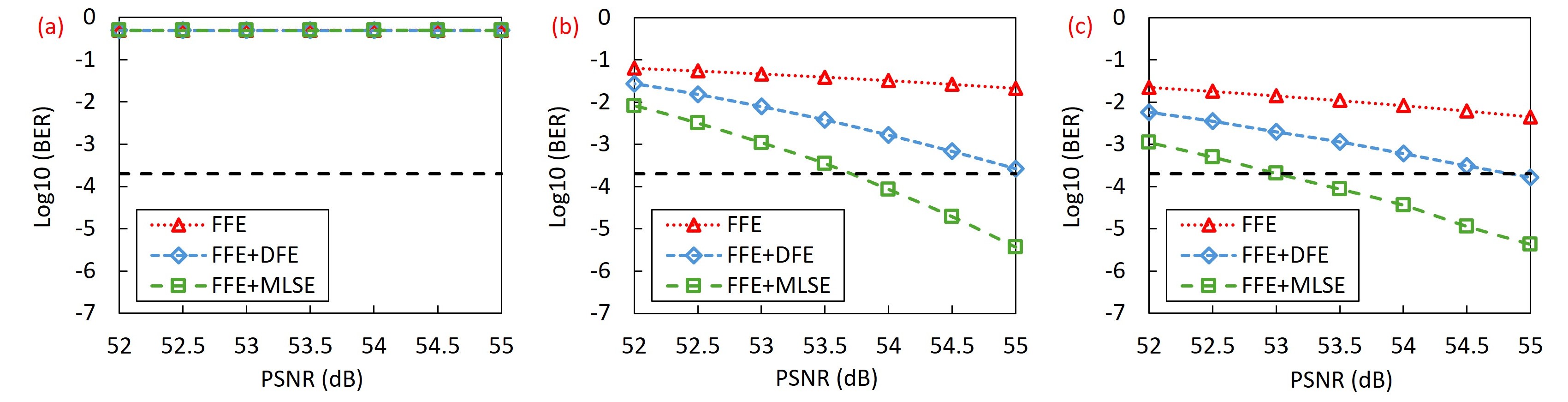}}
\caption{BER performance of (a) PAM4, (b) PAM6 based on cross 32QAM, and (c) PAM8 signals equalized by FFE, DFE, and MLSE using the channel 11. The black dashed line represents the KP4-FEC limit.
\label{Fig9}}
\end{figure*}

Fig.~\ref{Fig8}(a)-(c) shows the BER performance of PAM4, PAM6 based on framed-cross 32QAM, and PAM8 under the channel 1, respectively. For PAM4, DFE and MLSE satisfy the KP4-FEC limit at the required PSNRs of about 30.8 dB and 29 dB, respectively. For PAM6 based on framed-cross 32QAM, DFE and MLSE satisfy the KP4-FEC limit at the required PSNRs of about 30.6 dB and 30.03 dB, respectively. For PAM8, DFE and MLSE can achieve the KP4-FEC limit at the required PSNRs of about 33 dB and 32.5 dB, respectively. Fig.~\ref{Fig9}(a)-(c) shows the BER performance of PAM4, PAM6 based on cross 32QAM, and PAM8 over the channel 11, respectively. If only the FFE is adopted, none of the PAM4, PAM6, or PAM8 would satisfy the KP4-FEC limit. Due to the deep channel notches, PAM4 does not reach the adopted KP4-FEC limit with DFE or MLSE. The use of DFE enables PAM8 to meet the KP4-FEC limit at a PSNR of approximately 54.8 dB. With MLSE equalization, PAM6 and PAM8 reach the KP4-FEC limit at required PSNRs of approximately 53.7 dB and 53 dB, respectively.


\section{Conclusion}
\label{secV}
In this paper, we investigate PAM4/6/8 using FFE, DFE, and MLSE for a 400G-SerDes system. The approximate BER expressions for six types of PAM6 based on 32QAM templates are derived, which show that PAM6 based on framed-cross 32QAM achieves the lowest BER among the six PAM6 formats in the AWGN channel. In the AWGN channel, PAM6 and PAM8 exhibit PSNR penalties of about 4.2 dB and 7.2 dB at the KP4-FEC limit compared to PAM4, respectively. FFE, DFE, and MLSE are compared in the simulations of the 400G-SerDes system. After the channel 11 transmission, the PAM6 based on cross 32QAM with FFE and MLSE achieves the best BER performance among  PAM6 based on the six types of 32QAM templates, with a required PSNR of about 53.7 dB at the KP4-FEC limit. PAM6 based on framed-cross 32QAM with FFE and MLSE achieves the best BER performance under the channel 1 transmission among  PAM6 based on the six types of 32QAM templates, with a required PSNR of about 30.03 dB. Across most of the evaluated channels, PAM6 requires a lower PSNR than PAM4 and PAM8 to reach the KP4-FEC limit, demonstrating its potential for 400G-SerDes interconnects under a range of channel conditions. For channels with high insertion loss, PAM6 based on cross 32QAM with FFE and MLSE offers the best BER performance among the six PAM6 schemes.

\Acknowledgements{This work was supported by the National Key R\&D Program of China under Grant 2023YFB2905700, the National Natural Science Foundation of China under Grants 62371207 and 62005102, and the Young Elite Scientists Sponsorship Program by CAST under Grant 2023QNRC001.}






\bibliographystyle{scis}
\bibliography{reference}

\section*{Appendix A. Approximate BER Expressions for PAM6 Generated from 32QAM}

This appendix presents the approximate BER expressions for PAM6 generated from six types of 32QAM templates. For the Type $k$ 32QAM, each group of five input bits is mapped to one 32QAM constellation point in $\mathcal{S}_k$. $\mathcal{S}_k$ denotes the set of 32 constellation points obtained by removing four points from the 36QAM constellation with coordinate levels $(\pm 1, \pm 3, \pm 5)$. The in-phase and quadrature components of each 32QAM symbol are serialized to form a real-valued PAM6 sequence. The serialized PAM6 signal is impaired by AWGN with variance
\begin{equation}
\sigma^2=\frac{25}{\mathrm{PSNR}}
\end{equation}
where the peak PAM6 amplitude is normalized to 5. For a decision boundary with normalized distance $i$ from the transmitted level, the Gaussian-tail probability is

\begin{equation}
P_i
=
Q\left(\frac{i}{\sigma}\right)
=
Q\left(i\sqrt{\frac{\mathrm{PSNR}}{25}}\right).
\end{equation}

For each 32QAM template, the in-phase and quadrature bit-error events are analyzed separately. For the $\ell$th bit, where $\ell\in\{1,\ldots,5\}$, the in-phase analysis examines the bit-label transitions across all decision boundaries in each valid row of the constellation. The resulting in-phase bit-error probability can be expressed as

\begin{equation}
P_{\mathrm{b},I}^{(k,\ell)}
=
\sum_{i=1}^{9}
a_{k,\ell,i}
Q\left(i\sqrt{\frac{\mathrm{PSNR}}{25}}\right),
\end{equation}
where $a_{k,\ell, i}$ is obtained by counting the in-phase error events that change the $\ell$th bit at normalized distance $i$. Similarly, the quadrature bit-error probability is expressed as

\begin{equation}
P_{\mathrm{b},Q}^{(k,\ell)}
=
\sum_{i=1}^{9}
b_{k,\ell,i}
Q\left(i\sqrt{\frac{\mathrm{PSNR}}{25}}\right),
\end{equation}
where $b_{k,\ell, i}$ is determined from the quadrature bit-label transitions in each valid column. The in-phase and quadrature BER contributions are obtained by averaging over the five label bits:

\begin{equation}
P_{\mathrm{b},I}^{(k)}
=
\frac{1}{5}
\sum_{\ell=1}^{5}
P_{\mathrm{b},I}^{(k,\ell)},
\end{equation}

\begin{equation}
P_{\mathrm{b},Q}^{(k)}
=
\frac{1}{5}
\sum_{\ell=1}^{5}
P_{\mathrm{b},Q}^{(k,\ell)}.
\end{equation}
By combining the in-phase and quadrature BER contributions and using the product of the two contributions to account for their joint occurrence, the approximate BER is obtained as
\begin{equation}
P_{e,k} \approx
P_{\mathrm{b},I}^{(k)} + P_{\mathrm{b},Q}^{(k)} - P_{\mathrm{b},I}^{(k)}P_{\mathrm{b},Q}^{(k)}.
\end{equation}

\begin{table}[!t]
\caption{The coefficients of the approximate BER expression in Eq. (\ref{eq3}) for PAM6 based on the Type $k$ 32QAM.}
\label{Tab2}
\centering
\small
\renewcommand{\arraystretch}{1.3}
\setlength{\aboverulesep}{1pt} 
\setlength{\belowrulesep}{1pt} 
\resizebox{\textwidth}{!}{
\begin{tabular}{c c c c}
\midrule
$k$ & {$X_{k, i}$} & {$Y_{k, i}$} & {$Z_{k, i}$} \\
\midrule
1 & $\frac{3}{4}$, $0$, $\frac{9}{20}$, $0$, $\frac{1}{20}$, $0$, $\frac{1}{10}$, $0$, $-\frac{3}{20}$ 
& $\frac{13}{40}$, $0$, $\frac{1}{4}$, $0$, $\frac{1}{40}$, $0$, $0$, $0$, $-\frac{1}{20}$ 
& $\frac{17}{40}$, $0$, $\frac{1}{5}$, $0$, $\frac{1}{40}$, $0$, $\frac{1}{10}$, $0$, $-\frac{1}{10}$ \\
  
2 & 
\makecell{$\frac{13}{20}$, $\frac{1}{20}$, $\frac{3}{8}$, $\frac{1}{40}$, $\frac{1}{10}$, $\frac{1}{40}$, $\frac{1}{40}$, $-\frac{1}{40}$, $-\frac{1}{10}$} & \makecell{$\frac{13}{40}$, $\frac{1}{40}$, $\frac{3}{16}$, $\frac{1}{80}$, $\frac{1}{20}$, $\frac{1}{80}$, $\frac{1}{80}$, $-\frac{1}{80}$, $-\frac{1}{20}$} & \makecell{$\frac{13}{40}$, $\frac{1}{40}$, $\frac{3}{16}$, $\frac{1}{80}$, $\frac{1}{20}$, $\frac{1}{80}$, $\frac{1}{80}$, $-\frac{1}{80}$, $-\frac{1}{20}$} \\
        
3 & 
\makecell{$\frac{7}{10}$, $\frac{1}{10}$, $\frac{7}{20}$, $0$, $\frac{9}{40}$, $0$, $\frac{7}{40}$, $0$, $-\frac{1}{20}$} 
& \makecell{$\frac{3}{10}$, $\frac{1}{20}$, $\frac{1}{5}$, $0$, $\frac{1}{10}$, $0$, $\frac{3}{80}$, $0$, $0$} 
& \makecell{$\frac{2}{5}$, $\frac{1}{20}$, $\frac{3}{20}$, $0$, $\frac{1}{8}$, $0$, $\frac{11}{80}$, $0$, $-\frac{1}{20}$} \\
        
4 & 
\makecell{$\frac{11}{20}$, $\frac{1}{5}$, $\frac{2}{5}$, $\frac{1}{10}$, $\frac{3}{20}$, $0$, $0$, $-\frac{1}{10}$, $-\frac{1}{10}$} 
& \makecell{$\frac{11}{40}$, $\frac{1}{10}$, $\frac{1}{5}$, $\frac{1}{20}$, $\frac{3}{40}$, $0$, $0$, $-\frac{1}{20}$, $-\frac{1}{20}$} 
& \makecell{$\frac{11}{40}$, $\frac{1}{10}$, $\frac{1}{5}$, $\frac{1}{20}$, $\frac{3}{40}$, $0$, $0$, $-\frac{1}{20}$, $-\frac{1}{20}$} \\
        
5 & 
\makecell{$\frac{49}{80}$, $\frac{3}{20}$, $\frac{1}{4}$, $\frac{1}{10}$, $\frac{9}{80}$, $\frac{1}{40}$, $\frac{1}{20}$, $-\frac{3}{80}$, $-\frac{11}{80}$} 
& \makecell{$\frac{23}{80}$, $\frac{3}{40}$, $\frac{1}{8}$, $\frac{1}{20}$, $\frac{1}{8}$, $\frac{1}{40}$, $\frac{1}{40}$, $-\frac{1}{80}$, $-\frac{3}{40}$} 
& \makecell{$\frac{13}{40}$, $\frac{3}{40}$, $\frac{1}{8}$, $\frac{1}{20}$, $-\frac{1}{80}$, $0$, $\frac{1}{40}$, $-\frac{1}{40}$, $-\frac{1}{16}$} \\
        
6 & 
\makecell{$\frac{7}{10}$, $0$, $\frac{9}{20}$, $0$, $\frac{1}{20}$, $0$, $\frac{1}{20}$, $0$, $-\frac{3}{20}$} 
& \makecell{$\frac{3}{10}$, $0$, $\frac{9}{40}$, $0$, $\frac{3}{40}$, $0$, $\frac{1}{40}$, $0$, $-\frac{3}{40}$} 
& \makecell{$\frac{2}{5}$, $0$, $\frac{9}{40}$, $0$, $-\frac{1}{40}$, $0$, $\frac{1}{40}$, $0$, $-\frac{3}{40}$} \\
\bottomrule
\end{tabular}
}
\end{table}

The last term removes the overlap between in-phase and quadrature error events. After collecting terms with the same $Q$-function argument, the approximate BER expression can be written as
\begin{equation}
\begin{aligned}
P_{e,k} \approx \sum_{i=1}^9 X_{k, i} \cdot Q\left(i \sqrt{\frac{P S N R}{25}}\right) -
\left[\sum_{i=1}^9 Y_{k, i} \cdot Q\left(i \sqrt{\frac{P S N R}{25}}\right)\right] \cdot\left[\sum_{i=1}^9 Z_{k, i} \cdot Q\left(i \sqrt{\frac{P S N R}{25}}\right)\right]
\end{aligned}
\label{eq3}
\end{equation}
where $X_{k,i}$ represents the sum of the in-phase and quadrature bit-error coefficients, and $Y_{k,i}$ and $Z_{k,i}$ represent the corresponding in-phase and quadrature coefficients, respectively. The coefficients for PAM6 generated from the six types of 32QAM templates are listed in Table \ref{Tab2}.

\begin{figure}[!t]
\centerline{\includegraphics[width=6.5in]{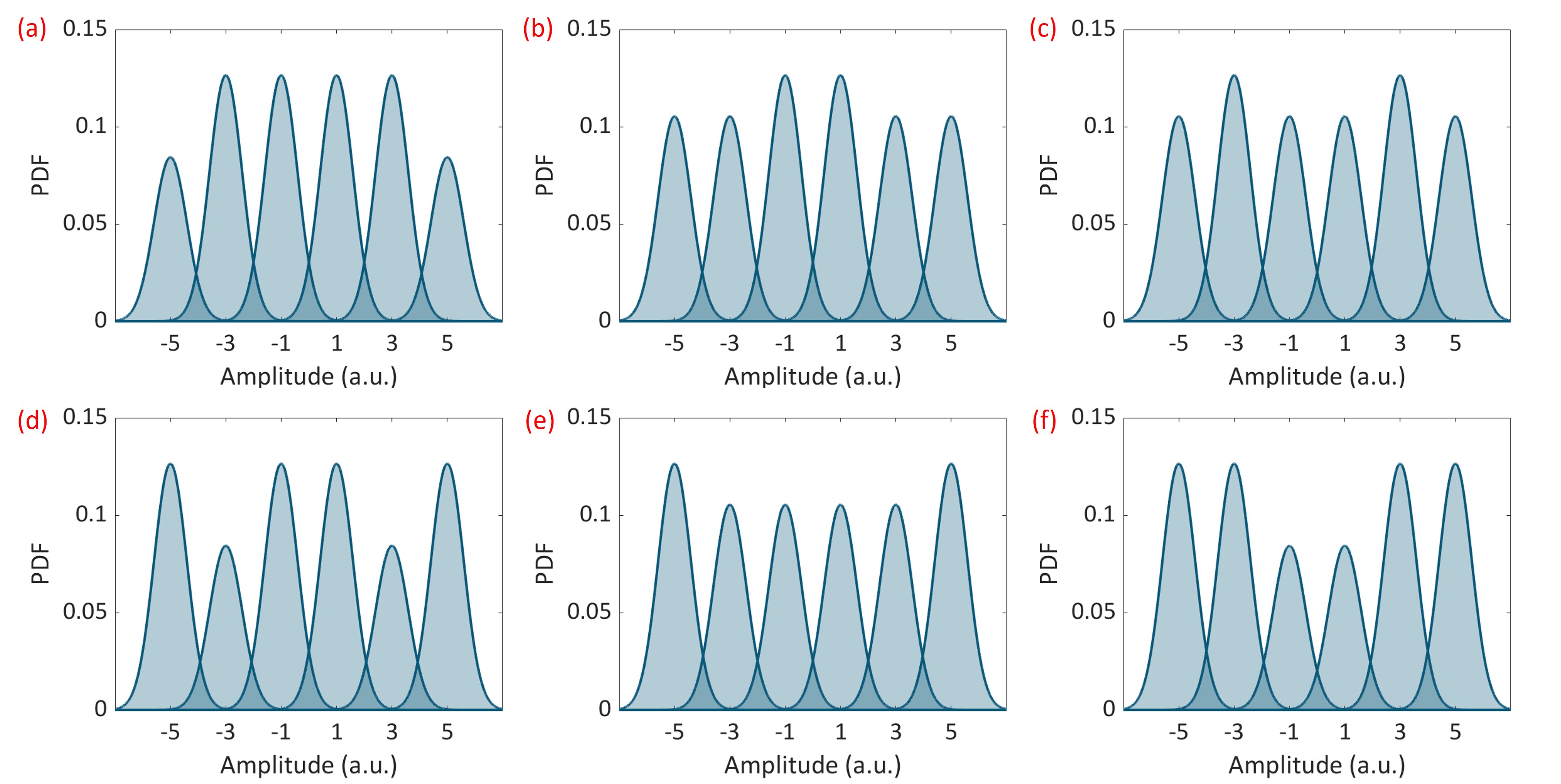}}
\caption{Probability density functions of the six 32QAM constellation templates used to generate PAM6 signals in the presence of additive Gaussian noise. (a) Type 1 (cross 32QAM), (b) Type 2, (c) Type 3, (d) Type 4 (framed-cross 32QAM), (e) Type 5, (f) Type 6, respectively.
\label{Fig10}}
\end{figure}

Fig.~\ref{Fig10} compares the probability density functions (PDFs) of PAM6 generated from the six types of 32QAM templates under an AWGN channel. The overlap between adjacent probability-density components around the decision thresholds characterizes the extent of the decision ambiguity regions. For the same peak amplitude and noise variance, PAM6 generated from framed-cross 32QAM exhibits the smallest total overlap between adjacent Gaussian components, thereby effectively reducing the probability of noise-induced symbol errors. Moreover, PAM6 generated from framed-cross 32QAM satisfies the Gray-mapping criterion. Consequently, the reduced decision ambiguity and Gray-mapping property jointly improve the BER, enabling PAM6 generated from framed-cross 32QAM to achieve the best BER performance among PAM6 based on the six types of 32QAM templates under a peak-power constrained AWGN channel.
 
\end{document}